\documentclass[notoc]{JHEP3}
\usepackage{epsfig}
\usepackage{amsmath}
\usepackage{amssymb}
\newcommand{\ie}{{\it i.e.}}
\newcommand{\eg}{{\it e.g.}}

\newcommand{\kvec}{\vect{k}}
\newcommand{\kt}{k_\perp}
\newcommand{\ellt}{\ell_\perp}
\newcommand{\pvec}{\vect{p}}

\newcommand{\vect}[1]{\boldsymbol{#1}}
\newcommand{\lqcd}{\Lambda_{QCD}}
\newcommand{\as}{\alpha_s}

\newcommand{\M}{{\cal M}}

\newcommand{\order}[1]{${\cal O}\left(#1 \right)$}

\newcommand{\eq}[1]{(\ref{#1})}

\newcommand{\beq}{\begin{equation}}
\newcommand{\eeq}{\end{equation}}
\newcommand{\nn}{\nonumber}
\newcommand{\beqa}{\begin{eqnarray}}
\newcommand{\eeqa}{\end{eqnarray}}
\newcommand{\be}{\begin{eqnarray}}
\newcommand{\ee}{\end{eqnarray}}
\newcommand{\beqat}{\begin{eqnarray*}}
\newcommand{\eeqat}{\end{eqnarray*}}

\newcommand{\inv}[1]{\frac{1}{#1}}

\title{\center{Semi-exclusive DVCS}}

\author{Paul Hoyer$^{1,2}$ and Heidi Virtanen$^{1}$\\
              $^1$Department of Physical Sciences and Helsinki Institute of
              Physics\\
              POB 64, FIN-00014 University of Helsinki, Finland \\
              $^2$NORDITA, Blegdamsvej 17, DK-2100 Copenhagen, Denmark\\
              E-mail: \email{paul.hoyer@helsinki.fi}, \email{heidi.m.virtanen@helsinki.fi}}
\preprint{\today\\ HIP-2006-31/TH \\ NORDITA-2006-19 \\  \hepph{0608281}}

\abstract{
We consider Semi-Exclusive Deeply Virtual Compton Scattering, $\gamma^*p \to \gamma Y$ (SECS), where $Y$ is an inclusive state of intermediate mass, $\lqcd \ll m_Y \ll Q$. When the photon is produced with a large transverse momentum $k_\perp \sim m_Y$ 
the subprocess is hard and the struck quark fragments independently of the target spectators. Using completeness this allows to express the SECS cross section in terms of ordinary parton distributions. Apart from direct comparisons with data (yet to come) new information on Bloom-Gilman duality may be obtained through comparisons of resonance production via DVCS ($\gamma^*p \to \gamma N^*$) with the SECS scaling distribution in $m_Y$.}


\begin{document}

\section{Semi-Exclusive Compton Scattering}
\label{sec:one}

Deeply Virtual Compton Scattering ($\gamma^* p \to \gamma p'$, DVCS) provides new insight into hadron structure \cite{Goeke:2001tz}. This process can be factorized into a hard, PQCD calculable subprocess ($\gamma^* q \to \gamma q'$, at lowest order) and a non-diagonal matrix element of the target, the Generalized Parton Distribution (GPD). The GPD probes novel aspects of the proton wave function since the final quark $q'$ of the subprocess joins the spectators to form the final proton with a momentum transfer $t$ from the target ($|t| \ll Q^2$, the virtuality of the photon).

DVCS brings two qualitatively new aspects compared to ordinary Deep Inelastic Scattering ($\gamma^* p \to X$, DIS):
\begin{itemize}

\item The hard subprocess is generalized from the vertex $\gamma^* q \to q$ to the process $\gamma^* q \to \gamma q'$. Thus DVCS provides a new hard probe for studying hadrons.

\item In DIS the struck quark hadronizes independently of the target spectators, whereas in DVCS the quark and the spectators combine coherently to form the final hadron.
\end{itemize}
It is only the second aspect which introduces the GPD. Here we wish to point out that there are processes which involve the new hard probe but can nevertheless be expressed using standard parton distributions.

The DVCS formalism applies also when the final proton is replaced with an $N^*$ resonance, as long as its mass is small compared to $Q$ and the CM energy $W$ of the $\gamma^* p$ system, $m_{N^*} \ll Q,\ W$. On the other hand, if instead of choosing a specific $N^*$ we allow the final quark $q'$ and the target spectators to form an inclusive  hadronic system $Y$ we may use completeness to relate the cross section to the discontinuity of a forward amplitude. We shall refer to this process as\footnote{A more accurate name would be Semi-Exclusive Deeply Virtual Compton Scattering. The abbreviation should not be confused with real Compton scattering.} Semi-Exclusive Compton Scattering ($\gamma^* p \to \gamma Y$, SECS). 

For the subprocess $\gamma^*q \to \gamma q'$ to be hard and thus perturbatively calculable the struck quark must be far off-shell after absorbing the $\gamma^*$. This implies that $x \neq x_B$, where $x$ is the fractional momentum of the struck quark and $x_B$ is the Bjorken variable\footnote{In DVCS the point $x=x_B$ is part of the hard process and gives rise to the imaginary part of the amplitude. In that case the singularity appears only in the loop integral over $x$, whereas in SECS it would be fixed by the external kinematics.}. The final photon will then have large transverse momentum $k_\perp$ (in the target rest frame, with the $\gamma^*$ momentum along the $z$-axis). In analogy to DIS, where $Q$ is of order $W$, it is natural to take $k_\perp$ of \order{m_Y}, with $m_Y$ the mass of the hadronic system $Y$.
When $k_\perp$ and $m_Y$ are large compared to the hadronic scale $\lqcd$ we may assume that the struck quark hadronizes independently of the spectators. As indicated in Fig.~1 the SECS cross section is then given by ordinary parton distributions. 

\EPSFIGURE[t]{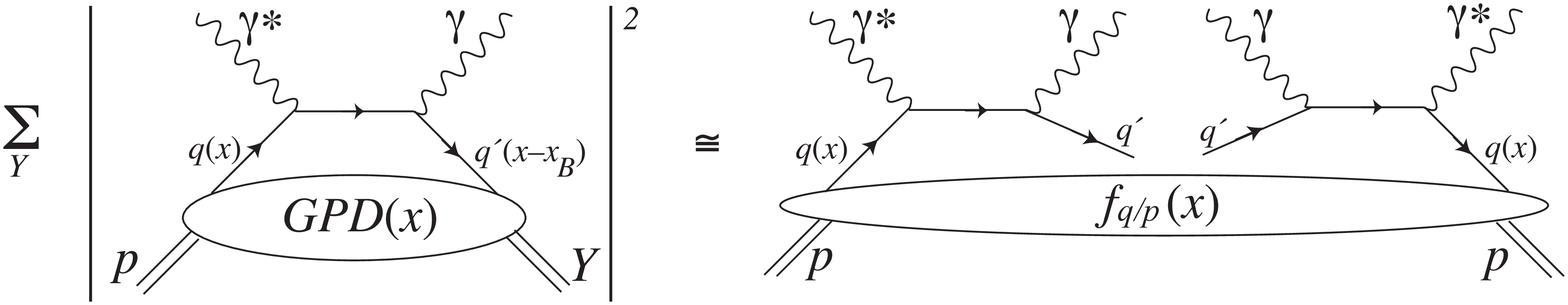,width=\columnwidth}{The cross section of Semi-Exclusive Compton Scattering $\gamma^* p \to \gamma Y$ (left) is given by a discontinuity of a forward amplitude. In the limit where the final quark $q'$ has large momentum it hadronizes independently of the spectators, and the cross section is given by the ordinary parton distributions $f_{q/p}(x)$ (right). In this case the value of the struck quark momentum fraction $x$ is fixed kinematically and given by \protect{\eq{xval}}.}

According to the above discussion, the kinematic region we shall consider for SECS is
\beq
W^2 \sim Q^2 \gg  m_Y^2 \sim \kt^2 \gg \lqcd^2 \label{secslim}
\eeq
Just as in the standard Bjorken limit of DIS and DVCS, $Q^2$ and $W^2$ are to be taken asymptotically large keeping the Bjorken variable
\beq \label{xbj}
x_B = \frac{Q^2}{Q^2+W^2}
\eeq
fixed. This ensures that the scattering occurs on a single parton in the target. The condition that the mass $m_Y$ of the inclusive hadronic system be much less than the total $\gamma^* p$ mass $W$ implies that the process is semi-exclusive, \ie, that there is a large rapidity gap between the hadrons and the photon in the final state. This is the kinematics of DVCS, which makes it plausible that the hard subprocess $\gamma^* q \to \gamma q'$ factorizes from the soft spectator dynamics.

Semi-exclusive processes have been previously considered \cite{Carlson:1993ys} for the scattering of real photons such as $\gamma p \to \gamma Y$, where the hard scale $Q^2$ of SECS is replaced with a large invariant momentum transfer $-t$ between the photons. Insofar as the subprocess $\gamma q \to \gamma q'$ can be factorized from the soft dynamics, arguments similar to the above ones may be invoked to express also this cross section in terms of standard parton distributions. In the absence of data on $\gamma p \to \gamma Y$ it was possible to roughly test this prediction \cite{Eden:2001ci} by relating it to elastic Compton scattering ($\gamma p \to \gamma p$) using Bloom-Gilman duality \cite{Melnitchouk:2005zr}. The calculation underestimated the measured cross section by an order of magnitude, which indicates that the assumed factorization into a hard subprocess and soft target dynamics does not hold even at large $-t$.

The failure of factorization for real, transverse photons may be related to the nonvanishing of the $\gamma \to q\bar q$ wave function in the `endpoint' region (where one quark carries vanishing momentum). This indeed prevents a QCD factorization proof for transverse photons in deeply virtual meson production \cite{Collins:1996fb}. There is independent experimental evidence \cite{Chekanov:2002rm} for a failure of factorization in $\gamma p \to \rho p$. The $\rho$ meson remains transversely polarized out to large momentum transfers, whereas quark helicity conservation would require longitudinal polarization \cite{Hoyer:2002qg}. By contrast, in the virtual photon process $\gamma^* p \to \rho p$ the $\rho$ does become longitudinally polarized with increasing virtuality $Q^2$ of the photon \cite{Breitweg:1998nh}, as required by factorization.

There are thus good reasons to believe that the subprocess $\gamma^* q \to \gamma q'$ may be factorized in SECS -- in analogy to factorization in DVCS. The difficulty in extracting GPD's from DVCS and related processes motivates checks of the validity of factorization for SECS. The apparent simplicity of SECS is somewhat offset by the requirement of the double limit in \eq{secslim}, which involves a hierarchy of two large scales. On the other hand, SECS also provides a new setting for testing Bloom-Gilman duality \cite{Melnitchouk:2005zr}. Does the (measured and predicted) cross section at large $m_Y$ describe also the resonance region ($Y=N,N^*$) in an average sense? This could teach us new aspects of GPD's, similarly as duality in DIS gives information on exclusive form factors.

\section{Kinematics and Cross Section}
\label{sec:two}

According to our discussion above, the conditions \eq{secslim} ensure that the deeply virtual process $ep \to e'\gamma Y$ occurs off a single parton, and that the final parton hadronizes independently of the target fragments. As illustrated in Fig.~1 this means that we can express the physical cross section using standard parton distributions\footnote{We omit the contribution of gluons, which is of \order{\as}.},
\beq\label{secsfact}
\frac{d\sigma}{dx}(ep \to e'\gamma Y) = \sum_q f_{q/p}(x)\,\hat\sigma(eq \to e'\gamma q')
\eeq
where $x$ is the fractional momentum carried by the struck quark. In the DVCS process (left hand side of Fig.~1, for a given hadron state $Y$) an integral over $x$ is implied, because only the overall momentum transfer from the target system to the subprocess is kinematically determined. In SECS the final quark hadronizes independently of the target and so the value of $x$ is fixed by kinematics (and given by \eq{xval} below).

\EPSFIGURE{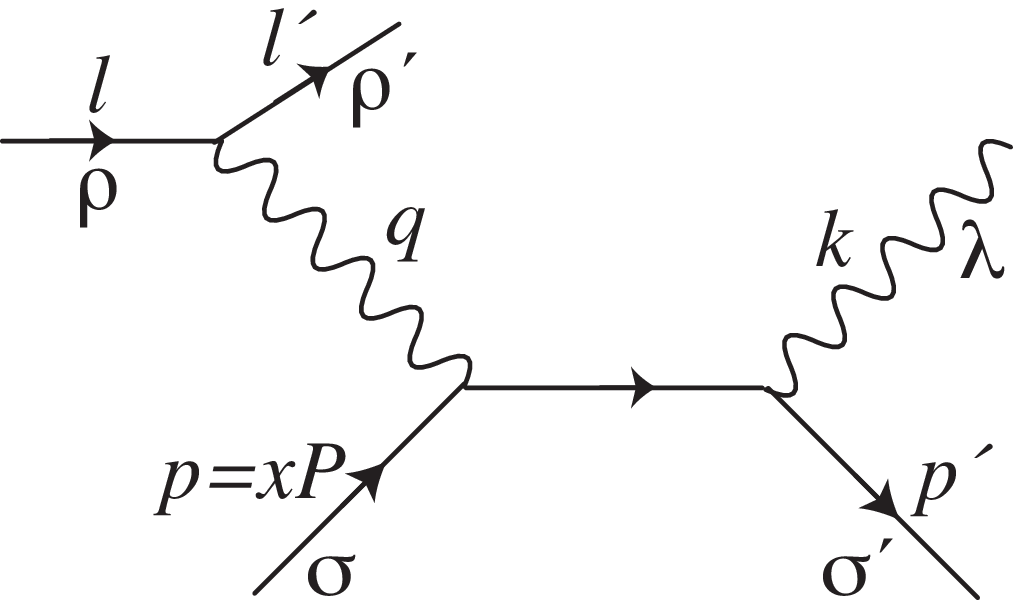,width=.3\columnwidth}{Notations for the momenta and helicities of the subprocess $eq \to e'\gamma q'$. The momentum of the target proton is denoted $P$.} 

Through a standard change of variables the SECS cross section can be written
\beq \label{sig1}
\frac{d\sigma}{dx_B\,dQ^2\,dx} = \frac{2f_{q/p}(x)}{(4\pi)^4} \frac{y^2}{xQ^2}
\int d^4 k\, \delta(k^2)\, \delta({p'}^2) \overline{|\M|^2}
\eeq
where $\M$ is the scattering amplitude for the subprocess $eq \to e'\gamma q'$ and the momentum assignments are shown in Fig.~2 ($p'=p+q-k$). The integral is Lorentz invariant and conveniently evaluated in a frame where the virtual photon momentum is along the negative $z$-axis. In the four-vector notation $p=(p^+,p^-,\pvec)$, where $p^\pm=p^0\pm p^3$ and $\pvec=(p^1,p^2)$ we have (neglecting masses)
\beqa \label{momdef}
\ell&=&\left(\frac{\ellt^2}{\ell^-},\ell^-,\vect{\ell}\right)\ ,\hspace{.7cm}
\ell'=\left(\frac{\ellt^2}{(1-y)\ell^-},(1-y)\ell^-,\vect{\ell}\right)\nn \\
p&=&(xP^+,0,\vect{0})\ ,\hspace{1cm} q=(-x_B P^+,y\ell^-,\vect{0})\ ,\hspace{2cm}
k=\left(\frac{\kt^2}{k^-},k^-,\kvec \right)
\eeqa
where $\ellt^2 \equiv \vect{\ell}^2$. We note the following relations,
\beqa
s &=& (\ell+P)^2 = \ell^- P^+\ ,\hspace{1cm}Q^2=-q^2\ ,\hspace{2cm} x_B=\frac{Q^2}{ys}\nn\\
\ellt^2 &=& \frac{1-y}{y^2}Q^2\ ,\hspace{2.4cm} {p'}^- = \frac{m_Y^2}{(1-x)P^+} 
= \frac{\kt^2}{(x-x_B)P^+}
\eeqa
where the last two equalities follow from ${p'}^2 = 0$ and $p_Y^2 = ((1-x)P+p')^2=m_Y^2$. They give
\beq \label{xval}
x=\frac{\kt^2+x_B m_Y^2}{\kt^2+m_Y^2}\ ,\hspace{2cm} x-x_B=\frac{(1-x_B)\kt^2}{\kt^2+m_Y^2}
\eeq

The scattering amplitudes $\M_{\sigma\sigma'}^{\rho\rho'}(\lambda)$ corresponding to Fig.~2 (and the related $u$-channel diagram) are, at lowest order and in the kinematic limits \eq{secslim},
\beqa
\M_{++}^{++}(\lambda) &=& -\M_{--}^{--}(-\lambda)= \frac{2\sqrt{2}e^3 e_q^2}{yQ}\left[(1-y)\sqrt{\frac{x-x_B}{x}}\,\delta_{\lambda,1}-\sqrt{\frac{x}{x-x_B}}\,\delta_{\lambda,-1}\right] \nn
\\ \\
\M_{--}^{++}(\lambda) &=& -\M_{++}^{--}(-\lambda)= \frac{2\sqrt{2}e^3 e_q^2}{yQ}\left[(1-y)\sqrt{\frac{x}{x-x_B}}\,\delta_{\lambda,1}-\sqrt{\frac{x-x_B}{x}}\,\delta_{\lambda,-1}\right] \nn
\eeqa
where we used the light-front spinors and polarization vectors of \cite{Brodsky:1997de}. All other helicity amplitudes vanish. The squared matrix element appearing in the cross section \eq{sig1} is then
\beq
\overline{|\M|^2} \equiv \inv{4}\sum_{\rho,\sigma,\lambda}\left|\M_{\sigma\sigma}^{\rho\rho}(\lambda)\right|^2 = \frac{4e^6e_q^4}{Q^2}\,\frac{1+(1-y)^2}{y^2}\,\left[\frac{x}{x-x_B}+\frac{x-x_B}{x}\right]
\eeq
With a change of variable $x \to m_Y^2$ the SECS cross section becomes
\beq \label{sig2}
\frac{d\sigma(ep\to e\gamma Y)}{dx_B\,dQ^2\,dm_Y^2\,d\kt^2} = f_{q/p}(x)\,\frac{\alpha^3\,e_q^4}{Q^6}\,[1+(1-y)^2]\,x_B(1-x_B)\,\left[\inv{(1-x_B)^2 \kt^2} + \frac{\kt^2}{(\kt^2+x_B m_Y^2)^2} \right]
\eeq
with $x$ given by \eq{xval}.

\section{Discussion}
\label{sec:three}

Semi-Exclusive Compton Scattering (SECS) provides a new tool for the study of hadron structure. Similarly to DVCS, the target hadron is probed by a hard subprocess, $\gamma^* q \to \gamma q'$, which is a generalization of the hard vertex $\gamma^* q \to q$ well-known from DIS. In the kinematics of \eq{secslim} the final quark $q'$ is fast compared to the target spectators and thus hadronizes independently. This implies a drastic simplification compared to DVCS, where $q'$ and the spectators coherently form the final hadron. The SECS cross section is given by standard parton distributions rather than by the Generalized Parton Distributions (GPD's) required for DVCS.

Similarly to deeply virtual exclusive processes, the semi-exclusive ones may be generalized to meson production, \eg, $\gamma^* p \to \pi Y$. The meson is produced in a compact configuration described by its distribution amplitude. Comparisons of the various semi-exclusive cross sections with data will test factorization and thus also check whether GPD's can be reliably extracted from data on exclusive processes. Quantitative comparisons will require to consider also the contribution from scattering on gluons, which was not included in the above analysis. The limit \eq{secslim} involves a hierarchy of two large scales ($m_Y \sim \kt \ll W \sim Q$), which nominally requires high energy collisions. 

Analogously to Bloom-Gilman duality in DIS \cite{Melnitchouk:2005zr}, one may ask whether data on resonance production in $\gamma^* p \to \gamma N^*$ agrees, in an average sense, with the smooth scaling curve measured (and calculated) for the SECS process $\gamma^* p \to \gamma Y$ at high $m_Y$. As seen from \eq{xval} the value of the parton momentum fraction $x$ depends only on the ratio $\kt/m_Y$ (at fixed $x_B$). $N^*$ production at relatively low $\kt$, which is given by the GPD's, may thus be directly compared with the high $m_Y$ continuum at a larger $\kt$, which is determined by ordinary parton distributions.

\acknowledgments
We are grateful to Olena Linnyk, who participated in the early stages of this work, and to Markus Diehl for valuable comments. The research was supported in part by the Academy of Finland through grant 102046.

\providecommand{\href}[2]{#2}\begingroup\raggedright
\endgroup

\end{document}